\documentclass[aps,prl,floats,twocolumn,epsf]{revtex4}
\usepackage{graphicx}
\usepackage{epsfig}

\def \beq{\begin{equation}}
\def \eeq{\end{equation}}
\def \beqarr{\begin{eqnarray}}
\def \eeqarr{\end{eqnarray}}
\def \bspt{\begin{split}}
\def \espt{\end{split}}
\def \bef{\begin{figure}}
\def \enf{\end{figure}}

\begin{document}

%%%%%%%%%%%%%%%%%%%%%%%%%%%%%%%%%%%%%%%%%%%%%%%%%%%%%%%%%%%%%%%%%%%%%

\title{Thermopower as a Possible Probe of Non-Abelian Quasiparticle Statistics in Fractional Quantum Hall Liquids}

\author{Kun Yang}

\affiliation{National High Magnetic Field Laboratoy
and Department of Physics, Florida State University, Tallahassee,
Florida 32306, USA}

\author{Bertrand I. Halperin}

\affiliation{Department of Physics, Harvard University, Cambridge, Massachusetts 02138}

\date{\today}

\begin{abstract}

We show in this paper that thermopower is enhanced in non-Abelian quantum Hall liquids under appropriate conditions. This is because thermopower measures entropy per electron in the clean limit, while the degeneracy and entropy associated with non-Abelian quasiparticles enhance entropy when they are present. Thus thermopower can potentially probe non-Abelian nature of the quasiparticles, and measure their quantum dimension.

\end{abstract}

\pacs{73.43.Cd}

\maketitle

Recently there has been very strong interest in unusual fractional quantum Hall (FQH) states\cite{day,collins}, whose quasiparticle excitations obey non-Abelian statistics\cite{nayak08}. Such interest is partially driven by the potential of using non-Abelian quasiparticles for quantum information storage and processing in an
intrinsically fault-tolerant
fashion~\cite{nayak08,kitaev03,freedman02,dassarma05,bonesteel05}. At this time the most promising candidate for non-Abelian statistics is the FQH state at filling factor $\nu_0=5/2$\cite{willett87}, in which the electrons in the half-filled first excited Landau level may condense into the Moore-Read (MR, or Pfaffian) state\cite{moore91} or its particle-hole conjugate (anti-Pfaffian state)\cite{lee07,levin07}, whose elementary quasiparticles have charge $e^*=e/4$.  Theoretical support for the Pffafian or anti-Pfaffian state as an explanation for the FQH state at $\nu_0=5/2$ has come from a variety of numerical calculations\cite{morf,rh,wan06,feiguin,peterson,moller}.

Phenomenologically, the 5/2 state looks very similar\cite{willett87,pan} to other FQH or integer quantum Hall states in ordinary transport measurements: one sees a quantized Hall resistance plateau and thermally activated longitudinal resistance. However, recent measurements, which involve tunneling between opposite edges across a constriction, have probed the quasiparticle charge $e^*$
 \cite{dolev,radu} and may have also seen effects of non-Abelian statistics
 \cite{willett}.
In this work we argue that bulk thermoelectric measurements, in particular thermopower, can also reveal the statistical properties of the non-Abelian quasiparticles under appropriate conditions. This is possible because thermopower can probe the entropy carried by non-Abelian quasiparticles, which is larger than that of Abelian quasiparticles  at low-temperature.

A key property of non-Abelian statistics is the appearance of ground state degeneracy $D$ that grows (up to an $O(1)$ prefactor) exponentially with the number of quasiparticles present in the system, $N_q$, when their positions are fixed:
\beq
D\sim d^{N_q},
\label{degeneracy}
\eeq
where $d > 1$ is the quantum dimension\cite{nayak08} of the quasiparticle. For the non-Abelian quasiparticles in the MR Pfaffian or anti-Pfaffian state, $d=\sqrt{2}$. We will use them as the primary examples of our discussion below, although essentially all of our discussions apply to other non-Abelian FQH states. Such degeneracy results in a ground state entropy
\beq
S_d=k_B\log D = k_B N_q\log d +O(1),
\label{entropy}
\eeq
where $k_B$ is the Bolzmann constant; {\em i.e.}, each quasiparticle carries entropy $k_B\log d$. In principle, there exists very weak coupling among the quasiparticles that can lift the ground state degeneracy\cite{anyonchain}; however such coupling vanishes exponentially as a function of the distance between quasiparticles. Thus the entropy formula Eq. (\ref{entropy}) remains valid as long as the temperature $T$ satisfies a condition
\beq
T_0\ll T \ll T_1,
\label{Trange}
\eeq
where $T_0\sim \Delta e^{-l/l_0}$ ($\Delta$ is quasiparticle gap, $l$ is the distance between quasiparticles and $l_0$ is the characteristic size of the quasiparticle) is the temperature scale associated with quasiparticle couplings, and $T_1$ is the temperature scale associated with other (ordinary) excitations in the system, including those related to the quasiparticles' positional degrees of freedom.
%below which their contribution to entropy goes to zero.
In principle, $T_0$ can be extremely low near the center of the quantum Hall plateau due to its exponential dependence on quasiparticle density, while $T_1$ should be larger. In particular, if the density of quasiparticles is sufficiently low, we expect that the quasiparticles will form a Wigner crystal due to the repulsion between quasiparticles, so the positional entropy should indeed disappear at low temperatures. We shall return to this issue later.

In a uniform system, the number of quasiparticles at low temperatures will be proportional to the deviation of the magnetic field $B$ from the value $B_0$ at the center of the plateau, where the filling fraction is equal to the ideal value $\nu_0$:
\beq
N_q=|(e/e^*)(B-B_0)/B_0|N_e,
\label{quasiparticlenumber}
\eeq
where $N_e$ is the number of electrons in the system. As a result the entropy $S_d=k_B N_q\log d$ grows linearly as $B$ deviates from the center of the plateau $B_0$, within the temperature range (\ref{Trange}).

This entropy due to the presence of non-Abelian quasiparticles can be probed using thermopower. In a thermopower measurement, one sets up a temperature gradient $\nabla T$, and voltage gradient ${\bf E}=-\nabla V$ is generated by the system to compensate for its effect so that no net electric current is flowing; the ratio between them,
\beq
Q=-\nabla V/\nabla T
\eeq
is the thermopower (also known as the Seebeck coefficient). It is well known\cite{chaikin} that under suitable circumstances, $Q$ measures the ``entropy per charge carrier" in the system. This has been rigorously justified for electrons in a strong magnetic field in the {\em clean} limit, first for non-interacting electrons\cite{obraztsov} and then for interacting electrons\cite{cooper}, so
\beq
Q=-S/(eN_e).
\label{entropyperparticle}
\eeq
In the following we present a derivation of (\ref{entropyperparticle}) that is slightly simpler than, but closely related to the arguments presented in Ref. \onlinecite{cooper}. For an electron liquid without impurity scattering,  the absence of net particle current requires that the variation of the liquid's internal pressure $P$ balances with external potential $\phi$:
\beq
\nabla P =({\partial P\over \partial \mu})_T\nabla \mu+({\partial P\over \partial T})_\mu \nabla T= -n\nabla \phi.
\label{pressure}
\eeq
Here $n=N_e/A$ is electron number density,  $A$ is area, and $\mu$ is the local chemical potential measured from $\phi$. The electrochemical potential is thus $\xi=\mu+\phi$, which is what an ideal voltage contact measures. From the grand potential relation
\beq
d\Omega=-SdT-PdA-N_ed\mu,
\eeq
follows the Maxwell relations $({\partial P\over \partial \mu})_{T,A}=({\partial N_e\over \partial A})_{T,\mu}=N_e/A=n$ and $({\partial P\over \partial T})_{\mu,A}=({\partial S\over \partial A})_{T,\mu}=S/A$. The last steps follow from the extensiveness of $S$, $N_e$, and $A$, which are proportional to each other when intensive quantities $\mu$ and $T$ are fixed. Thus we find
\beq
n\nabla\mu+(S/A) \nabla T=-n\nabla\phi,
\eeq
or
\beq
\nabla \xi/\nabla T=-S/N_e.
\label{SNe}
\eeq
The voltage measured by voltmeter with ideal contacts is $\Delta\xi/q$, where $q$ is the charge of the liquid's constituent particle, for electrons $q=-e$ while for holes $q=e$. Thus Eq. (\ref{entropyperparticle}) follows for electron samples; for hole samples there is a corresponding sign change.

The simplicity of the argument above suggests the result (\ref{entropyperparticle}) applies even in the {\em absence} of magnetic field, in the clean limit. We note that when studying thermoelectric effects, one usually starts with transport equations\cite{mahan}, and thermopower is expressed as a ratio between transport coefficients\cite{mahan,cooper,note}. In the absence of both disorder {\em and} magnetic field, transport coefficients are divergent and not well-defined; however thermopower is still well-defined and finite, and can be obtained easily using the hydrodynamic arguments presented above.

Strictly speaking, the hydrodynamic analysis above applies to a liquid whose internal stress tensor has only a diagonal component $P$. When the quasiparticles form a Wigner crystal, it may sustain some shear stress when driven out of equilibrium; this may result in correction to Eq. (\ref{pressure}), which is proportional to the product of shear strain gradient (if present) and shear modulus of the crystal. However due to the long-range nature of the Coulomb interaction and the very small percentage of charge that actually form the crystal, we expect the shear modulus to be much smaller than the bulk modulus, and such correction should be negligible.

Combining Eqs. (\ref{entropy},\ref{quasiparticlenumber},\ref{entropyperparticle}) we find within the temperature window (\ref{Trange}) and in the clean limit,
\beq
Q=-|(B-B_0)/B_0|(k_B/|e^*|)\log d.
\label{thermopower}
\eeq
Since $|e^*|$ can be measured independently\cite{dolev,radu,willett}, Eq. (\ref{thermopower}) suggests that thermopower gives a direct measurement of quantum dimension $d$ in the clean limit. It should be emphasized that it is $d > 1$ that directly reveals the non-Abelian nature of the quasiparticle, while a fractional charge may correspond to either Abelian or non-Abelian quasiparticles.  We note that in
the low-temperature regime  we are discussing here, phonons will be frozen out so that extrinsic effects like phonon drag are absent; thus thermopower should probe the intrinsic properties of the electron system.

We now turn the discussion to the temperature range (\ref{Trange}) within which our entropy formula (\ref{entropy}) is valid.
%especially the scale of $T_1$ below which we can neglect the positional entropy of the quasiparticles.
If the quasiparticles form a Wigner crystal, positional entropy comes from magnetophonons at low $T$, and one would expect
$T_1 \approx T_D$, where $T_D$ is the maximum phonon energy or Debye temperature. Treating the quasipartciles as point particles with charge $e^*$ moving in the magnetic field $B$, they form a triangular lattice with lattice spacing
\beq
a=l_B\left[{4\pi\over \sqrt{3}\nu_0}{e\over e^*}{B_0\over |B-B_0|}\right]^{1/2},
\eeq
where $l_B$ is the magnetic length. Using the known magnetophonon spectrum of that system\cite{fukuyama}, we obtain
\beq
k_BT_D \sim {e^2\over\epsilon l_B}\sqrt{{e\over |e^*|}}\left[{\nu_0|B-B_0|\over B_0}\right]^{3\over 2},
\eeq
To justify treating quasiparticles as real particles for the specific case of $\nu_0=5/2$, we observe that they are vortices of a paired composite fermion superconductor; using a duality transformation these vortices become particles, and the background composite fermion Cooper pairs become a magnetic field. While the short-range part of the quasiparticle interaction is not known, the long-range part is determined by the charge $e^*$, which is the most important in the low-density limit.

Another important temperature here is the melting temperature $T_m$. Its classical value is a small fraction of the Coulomb interaction energy between quasiparticles:
\beq
k_BT_m = {1\over \Gamma}{(e^*)^2\over\epsilon l_B}\left[{\nu_0|B-B_0|\over 2B_0}{e\over |e^*|}\right]^{1\over 2},
\label{freezing}
\eeq
where $\Gamma\approx 137$\cite{grimes,gann,morf79}. Thus $T_m$ and $T_D$ have {\em different} dependences on $B-B_0$. This allows for the interesting possibility of $T_m < T_D$. If melting is continuous or very weakly 1st order, the liquid state that results from melting is expected to have strong short-range crystal order, and its positional entropy remains to be small compared to $S_d$ as along as $T \ll T_D$, as a result we expect $T_m < T_1 < T_D$ in this case. On the other hand if melting is a strong 1st order transition with latent heat of order $k_BT_m$ per quasiparticle, then we have $T_1=T_m$.

For highest quality samples where the $5/2$ FQH plateaus are observed, we typically have $B_0\approx 4T$ which results in $l_B\approx 100{\AA}$, and at the edge of plateau $|B-B_0|/B_0\approx 1/200$, indicating the quasiparticles form a (pinned) Wigner crystal up to that point, at low temperature. Using the dielectric constant $\epsilon=13$ and $e^*/e=1/4$, we obtain $T_m\approx 7mK$ and $T_D\approx 300mK$ at 5/2 plateau edge. We indeed have $T_m \ll T_D$ in this case.

To estimate $T_0$, we choose $l_0=\sqrt{|e/e^*|}l_B$ which is the quasiparticle magnetic length, and $l=a$. Combining with $\Delta\approx 0.5K$ we obtain $T_0 \lesssim 1mK $ on the plateau. We note these estimates are quite rough, especially that of $T_0$, due to the uncertainty in $l_0$ which enters the exponential.

In general, the presence of disorder will give corrections to the result (\ref{entropyperparticle}). In particular, a quasiparticle Wigner crystal is expected to be pinned by {\em weak} disorder in the linear response regime, which is what gives rise to the FQH plateau in the first place. Pinning will also suppress its contribution to thermopower. Thus in order to observe the predicted effect on thermopower, one needs to de-pin the quasiparticles. The most straightforward way to do that is to melt the quasiparticle Wigner crystal by having $T > T_m$. To ensure positional entropy being small compared to $S_d$, we need $ T \ll T_D$, {\em and} melting being a continuous or weak 1st order transition. Experiment\cite{grimes} as well as numerical simulation of classical Coulomb system suggest this is indeed the case\cite{gann,morf79,he,kalia}. For $T\gtrsim T_m$, the liquid has strong short-range crystal order, and positional entropy can be estimated by summing the contributions from magnetophonons, and free dislocations (which triggers melting in 2D). Just like in the crystal phase, the phonon contribution is small compared to $S_d$ as long as $T \ll T_D$. The dislocation contribution
\beq
S_{dis} \approx N_{dis}\log (N_q/N_{dis}),
\eeq
where $N_{dis}$ is the number of free dislocations in the system. Thus $S_{dis} \ll S_d$ as long as $N_{dis} \ll N_q$. At low-T we expect $N_{dis}/N_q \sim e^{-E_c/k_BT}$, where $E_c$ is the dislocation core energy. Using results from a classical calculation at $T=0$ \cite{fisher}, one finds
\beq
E_c\approx 0.11 {(e^*)^2\over\epsilon l_B}\left[{\nu_0|B-B_0|\over 2\pi B_0}{e\over |e^*|}\right]^{1\over 2},
\eeq
or $E_c/k_BT_m\approx 8$. One needs to caution here though both quantum and thermal fluctuations can renormalize $E_c$ downward\cite{note1}.

While disorder cannot pin a quasiparticle liquid, it can still give rise to significant resistance as a liquid with a low density of dislocations tend to be very viscous. Thus in order to observe the non-Abelian entropy through Eqs. ({\ref{thermopower}), we need to work in the temperature range
\beq
T_m\lesssim T \ll T_D,
\label{newTrange}
\eeq
{\em and} with sufficiently clean sample. The sample should be clean enough such that within the range of Eq. (\ref{newTrange}), the Hall resistivity $\rho_{xy}$ is close to its classical value reached at high temperature, while the longitudinal resistivity $\rho_{xx}$ is small compared to the quasiparticle contribution to $\rho_{xy}$.

Throughout our analysis, we have assumed that variations in $\nu$ due to inhomogeneities in the electron density are small compared to the average value of $|\nu-\nu_0|$, which puts additional stringent condition on sample quality.  We have also assumed that there is not a short-range attraction between quasiparticles strong enough to overcome their Coulomb repulsion and cause binding between pairs.  If binding occurs, then quasiparticles might form a Wigner crystal of charge $e/2$ pairs, for small values of $|B-B_0|$, and the entropy $S_d$ would be lost.

Another possible concern is that since the non-Abelian degeneracy (\ref{degeneracy}) is {\em not} associated with individual quasiparticles, but is a global property, the system might have difficulty accessing all the (nearly) degenerate states and the associated entropy at very low temperature. We do not believe this will be a problem in a thermopower measurement. Thermopower is driven, physically, by effects at the edges of the sample, where equilibrium is established between electrons in a lead or contact and quasiparticles within the two-dimensional quantized Hall system. This necessarily assumes that there is some reasonable rate of hopping of charge back and forth between the two-dimensional system and the leads, with creation and annihilation of quasiparticles close to the edge.  As a result of this hopping, there should be a considerable amount of braiding in the edge region, which should give access to the full entropy of the states near the edges. We believe this is all that is required for the entropy to show up in a measurement of thermopower. However, we expect that even in the bulk, in the Wigner crystal
phase, there will be significant braiding of quasiparticles on the laboratory time scale, due to motion of dislocations, interstitials, and vacancies. Moreover, even if one neglects braiding, splitting of the ground state degeneracy due to the exponentially small interactions between quasiparticles will still correspond to a rate that is fast on a laboratory time scale, if one is not extremely close to the center of the plateau in a very uniform sample.

Thermopower has been studied in 2D electron gas in a magnetic field (especially in the quantum Hall regimes), both theoretically\cite{girvin,cooper} and experimentally \cite{obloh,ying}. Experimentally it was found that $Q$ reaches minima as a function of magnetic field on integer and fractional quantum Hall plateaus, and vanishes (apparently) exponentially as $T\rightarrow 0$ there.
Thermopower is bigger at filling factors corresponding to compressible states, but still vanishes as $T\rightarrow 0$, typically in a power-law manner\cite{ying}. The central result of this work is that thermopower can be strongly enhanced near filling factors where a non-Abelian quantum Hall state is realized, and takes a roughly temperature-independent value within the temperature range (\ref{newTrange}), that depends on the quantum dimension of the non-Abelian quaisparticle in sufficiently clean samples.

The mechanism for thermopower enhancement discussed here also applies to entropy generated by more conventional source of degeneracy, like electron spin. Specific examples include the Wigner crystals formed on the integer quantum Hall plateaus around $\nu=2n$, where $n$ is an integer. In this case the quasiparticles are simply electrons or holes, and if the Lande $g$-factor is tuned to be very close to zero by applying proper pressure, they each carry a spin entropy $k_B\log 2$ for temperature above the very small Zeeman splitting. As a result Eq. (\ref{thermopower}) applies in the appropriate temperature range, with $|e^*|=e$ and $d=2$. There are several advantages in attempting to observe the physics discussed here in these systems, as compared to the non-Abelian FQH states. (1) The gap is bigger and quantized plateau wider, allowing for a bigger field range for exploration. (2) Combined with bigger quasiparticle charge, this leads to higher $T_D$ and $T_m$; these lead to a more accessible and possibly wider range of temperature for the validity of Eq. (\ref{thermopower}).

We are indebted to Gabor Csathy and Jim Eisenstein for discussions that led to the present work. We also benefited from very useful discussions with and comments from Nick Bonesteel, Chetan Nayak, Nick Read, Gil Refael, and Ady Stern. This work was supported in part by NSF grants DMR-0704133 (KY)
and DMR-0541988 (BIH), and by the Microsoft Corporation (BIH).

{\em Note added} -- The present paper supersedes an earlier manuscript\cite{yang} by one of us on the same subject. Very recently a new preprint\cite{cooperstern} appeared, in which the authors use ideas closely related to those discussed here to explore possibilities of probing non-Abelian entropy under equilibrium situations.


\begin{thebibliography}{99}
\bibitem{note} In the presence of a magnetic field, transport coefficients are matrices with both longitudinal and Hall components; in the absence of disorder only Hall components are non-zero.
\bibitem{note1} Another potential concern is once melted, quasiparticles are mobile and can circle (or braid) around each other; such braiding can lift the non-Abelian degeneracy $D$. However as long is the rate of such braiding is small compared to $k_BT/\hbar$, the non-Abelian entropy $S_d$ remains, which is most likely the case in the temperature range of interest (see Eq. (\ref{newTrange}) below).
\end{thebibliography}
\end{document}